\documentclass[ showpacs,preprintnumbers,amsmath,amssymb,floatfix,prd]{revtex4}

\usepackage{graphicx}
\usepackage{dcolumn}
\usepackage{bm}

\newcommand{\beq}{\begin{equation}}
\newcommand{\eeq}{\end{equation}}
\newcommand{\bey}{\begin{eqnarray}}
\newcommand{\eey}{\end{eqnarray}}

\begin{document}

\title{
Emergence and Expansion of Cosmic Space in BIonic system
 }

\author{A. Sepehri}
\email{ alireza.sepehri@uk.ac.ir} \affiliation{Faculty of Physics,
Shahid Bahonar University, P.O. Box 76175, Kerman, Iran}
\author{Farook Rahaman}
\email{rahaman@iucaa.ernet.in} \affiliation{Department of
Mathematics, Jadavpur University, Kolkata 700032, West Bengal,
India}
\author{Anirudh Pradhan}
\email{ pradhan@iucaa.ernet.in} \affiliation{Department of
Mathematics, GLA University, Mathura-281 406, U.P., India.}

\author{Iftikar Hossain Sardar}
\email{iftikar.spm@gmail.com} \affiliation{Dept.  of Mathematics,
Jadavpur University, Kolkata 700032, West Bengal, India}

\begin{abstract}
Recently, Padmanabhan [arXiv:1206.4916] argued that the expansion
rate of the universe can be thought of as the emergence of space
as cosmic time progresses and is related to the difference between
the surface degrees of freedom on the holographic horizon and the
bulk degrees of freedom inside. The main question arises as to
what is origin of emergence of space in 4D universe? We answer to
this question in BIonic system. The BIon is a configuration in
flat space of a D-brane and a parallel anti-D-brane connected by
 a thin shell   wormhole with F-string charge. We propose
a new model that allows all degrees of freedom inside and outside
the universe are controlled by the evolutions of BIon in extra
dimension and tend to degrees of freedom of black F-string in
string theory or black M2-brane in M theory.

\end{abstract}

 \maketitle
\section{Introduction}
Intrigued by the holographic principle, Padmanabhan recently
proposed a novel idea, saying that our cosmic space is emergent as
cosmic time progresses. The emergence is governed by the basic
relation that the increase rate of Hubble volume is linearly
determined by the difference between the number of degrees of
freedom on the horizon surface and the one in the bulk \cite{f1}.
Until now, this idea has been discussed in many papers \cite{f2,
f3, f4, f5, f6, f7}. For example, , following Padmanabhan
proposal, some authors generalized the basic relation to derive
the Friedmann equations of an (n + 1)-dimensional
Friedmann-Robertson-Walker universe corresponding to general
relativity, Gauss-Bonnet gravity, and Lovelock gravity \cite{f2}.
Some other authors generalized Padmanabhan paradigm to brane
world, scalar-tensor gravity, and f(R) theory, respectively, and
found that in the cosmological setting the Friedmann equations can
be successfully derived \cite{f3}. In another scenario, by
modification Padmanabhan idea, the Friedmann equation of the
Friedmann-Robertson-Walker (FRW) Universe with any spatial
curvature was derived and  the study to higher dimensional
spacetime is extended \cite{f4}. In another paper , researchers
applied this method to a non flat universe, and modified the
evolution equation to lead to the Friedmann equation \cite{f5}. On
the other hand, some investigators discussed that Padmanabhan's
conjecture holds for the flat Friedmann-Robertson-Walker universe
in Einstein gravity but does not hold for the non-flat case unless
one uses the aerial volume instead of the proper volume \cite{f6}
and in more recent investigations, using emergence of cosmic space
framework and by considering a generic form of the entropy as a
function of area, the general dynamical equation of FRW universe
filled with a perfect fluid was derived \cite{f7}.

Now, the  main question arises as to what is origin of emergence
of space in 4D universe? We answer to this question in BIonic
system.  It might be reasonable to ignore the wormhole in the
ultraviolet where the branes and antibranes are well separated and
the brane's spike is far from the antibrane's spike, it is likely
that one BIon forms and grows where the spikes of brane and
antibrane meet each other \cite{f8,f9}. In this condition, there
exists many channels for flowing energy from extra dimensions into
our universe and as a result, degrees of freedom inside the
universe increase and tend to degrees of freedom of black F-string
in string theory or black M2-brane in M-theory.

The outline of the paper is as the following.  In section
\ref{o1}, we construct Padmanabhan proposal in ten dimensional
BIonic system and obtain degrees of freedom inside and outside of
universe in terms of wormhole parameters in extra dimension. In
section \ref{o2}, we consider Padmanabhan idea in eleven
dimensional M2-M5-brane. The last section is devoted to summary
and conclusion.

\section{ The Padmanabhan idea in ten dimensional BIonic system}\label{o1}
In this section, we will search the role of wormholes in
emergency  of cosmic space  and show that they are the main causes
of expansion of universe. To this end, we will follow
Padmanabhan's approach  in thermal BIon and discuss that the surface degrees
of freedom on the holographic horizon and the bulk degrees of
freedom inside the universe   depend on the temperature of BIon,
number of branes and the distance between branes.

 To
illustrate  the BIon we focus  to an embedding of the D3-brane
world volume in 10D Minkowski space-time having  line element:
\begin{eqnarray}
&& ds^{2} = -dt^{2} + dr^{2} + r^{2}(d\theta^{2} + sin^{2}\theta d\phi^{2}) + \sum_{i=1}^{6}dx_{i}^{2}.
\label{Q1}
\end{eqnarray}
without background fluxes. We choose    the world volume coordinates
of the D3-brane as $\lbrace\sigma^{a}, a=0..3\rbrace$ and  also define
$\tau = \sigma^{0},\,\sigma=\sigma^{1}$ as  the embedding of the
three-brane which  is given by \cite{f8,f9}:
\begin{eqnarray}
t(\sigma^{a}) = \tau,\,r(\sigma^{a})=\sigma,\,x_{1}(\sigma^{a})=z(\sigma),\,\theta(\sigma^{a})=\sigma^{2},\,\phi(\sigma^{a})=\sigma^{3}
\label{Q2}
\end{eqnarray}
and the remaining coordinates $x_{i}=2, ... , 6$ are constant. Note that here   only  one non-trivial embedding
function $z(\sigma)$ that expresses  the bending of the brane. Assume that  z be a
transverse coordinate to the branes and $\sigma$ be the radius on the world-volume.
The flat branes is thus defined as $z(\sigma) = 0$.
 Then the induced metric on the branes will get the form as
\begin{eqnarray}
\gamma_{ab}d\sigma^{a}d\sigma^{b} = -d\tau^{2} + (1 + z'(\sigma)^{2})d\sigma^{2} + \sigma^{2}(d\theta^{2} + sin^{2}\theta d\phi^{2})
\label{Q3}
\end{eqnarray}
so that the spatial volume element is $dV_{3}=\sqrt{1 +
z'(\sigma)^{2}}\sigma^{2}d\Omega_{2}$.
\\

 Note that we have assumed the universe in brane-antibrane system +
wormhole that connect these branes. Therefore, the shape function
comprising  the induced metric assumes the form \cite{MT}
\[ b(\sigma) = \frac{\sigma  {z^\prime}^2 }{1+{z^\prime}^2}\]
Here, we demand    two boundary conditions  as  $z(\sigma)\rightarrow
0$ for $\sigma\rightarrow \infty$ and $z'(\sigma)\rightarrow
-\infty$ for $\sigma\rightarrow \sigma_{0}$, where $\sigma_{0}$
is the minimal two-sphere radius of the configuration
 i.e. throat radius.  Now, we use   k units of F-string
charge along the radial direction to yield  \cite{f8,f9}:
\begin{eqnarray}
z(\sigma)= \int_{\sigma}^{\infty} d\acute{\sigma}\left(\frac{F(\acute{\sigma})^{2}}{F(\sigma_{0})^{2}}-1\right)^{-\frac{1}{2}}
\label{Q4}
\end{eqnarray}
BIon $F(\sigma)$ in finite temperature can be expressed as
\begin{eqnarray}
F(\sigma) = \sigma^{2}\frac{4\cosh^{2}\alpha - 3}{\cosh^{4}\alpha}
\label{Q5}
\end{eqnarray}
where $\cosh\alpha$ is defined as
\begin{eqnarray}
\cosh^{2}\alpha = \frac{3}{2}\frac{\cos\frac{\delta}{3} + \sqrt{3}\sin\frac{\delta}{3}}{\cos\delta}
\label{Q6}
\end{eqnarray}
with the definitions:
\begin{eqnarray}
\cos\delta \equiv \overline{T}^{4}\sqrt{1 + \frac{k^{2}}{\sigma^{4}}},\, \overline{T} \equiv \left(\frac{9\pi^{2}N}{4\sqrt{3}T_{D_{3}}}\right)T, \, \kappa \equiv \frac{k T_{F1}}{4\pi T_{D_{3}}}
\label{Q7}
\end{eqnarray}
In   equation (7), T and N   are    the finite temperature of BIon and number of D3-branes respectively.  $T_{D_{3}}$ and $T_{F1}$ are tensions of brane and
 fundamental strings respectively. We will construct wormhole configurations by  attaching a mirror solution to Eq. (\ref{Q4}).
  To measure the  separation distance $\Delta = 2z(\sigma_{0})$ between the N D3-branes
and N anti D3-branes for a given brane-antibrane wormhole configuration, one has to  define the four
parameters N, k, T and $\sigma_{0}$.

Now, we have:
\begin{eqnarray}
\Delta = 2z(\sigma_{0})= 2\int_{\sigma_{0}}^{\infty} d\acute{\sigma}\left(\frac{F(\acute{\sigma})^{2}}{F(\sigma_{0})^{2}}-1\right)^{-\frac{1}{2}}
\label{Q8}
\end{eqnarray}
For small temperature limit, we obtain:
\begin{eqnarray}
\Delta = \frac{2\sqrt{\pi}\Gamma(5/4)}{\Gamma(3/4)}\sigma_{0}\left(1 + \frac{8}{27}\frac{k^{2}}{\sigma_{0}^{4}}\overline{T}^{8}\right)
\label{Q9}
\end{eqnarray}

  Let us now construct the Padmanabhan idea in thermal BIon. For this, we
need to compute the contribution of the BIonic system to the
degrees of  the surface degrees of freedom on the holographic
horizon and the bulk degrees of freedom inside the universe. To
this end, we write the following relations between these degrees
of freedom and the entropy of BIon and also, the mass density
along the transverse direction,
 \begin{eqnarray}
&&N_{sur} + N_{bulk} = N_{BIon}= N_{brane} + N_{anti-brane} +
N_{wormhole}\nonumber \\&& \simeq
4L_{P}^{2}S_{BIon}=\frac{4T_{D3}^{2}}{\pi T^{4}}\int
d\sigma\frac{F(\sigma)}{\sqrt{F^{2}(\sigma)-F^{2}(\sigma_{0})}}\sigma^{2}\frac{4}{cosh^{4}\alpha}
\nonumber \\&& N_{sur} - N_{bulk} \simeq \int d\sigma
\frac{dM_{BIon}}{dz}=\frac{2T_{D3}^{2}}{\pi T^{4}}\int d\sigma
\frac{F(\sigma)}{F(\sigma_{0})}\sigma^{2}\frac{4cosh^{2}\alpha +
1}{cosh^{4}\alpha} \label{Q10}
\end{eqnarray}
Solving these equations simultaneously, we obtain:
\begin{eqnarray}
&&N_{sur} \simeq \frac{4T_{D3}^{2}}{\pi T^{4}}\int
d\sigma\frac{F(\sigma)}{\sqrt{F^{2}(\sigma)-F^{2}(\sigma_{0})}}\sigma^{2}\frac{4}{cosh^{4}\alpha}
+ \frac{2T_{D3}^{2}}{\pi T^{4}}\int d\sigma
\frac{F(\sigma)}{F(\sigma_{0})}\sigma^{2}\frac{4cosh^{2}\alpha +
1}{cosh^{4}\alpha} \nonumber \\&&  N_{bulk} \simeq
\frac{4T_{D3}^{2}}{\pi T^{4}}\int
d\sigma\frac{F(\sigma)}{\sqrt{F^{2}(\sigma)-F^{2}(\sigma_{0})}}\sigma^{2}\frac{4}{cosh^{4}\alpha}
-\frac{2T_{D3}^{2}}{\pi T^{4}}\int d\sigma
\frac{F(\sigma)}{F(\sigma_{0})}\sigma^{2}\frac{4cosh^{2}\alpha +
1}{cosh^{4}\alpha} \label{Q11}
\end{eqnarray}

This equation indicates  that with the  increase of  temperature in BIonic
system, the bulk degrees of freedom inside the universe will  increase.
A possible reason for this is
  when spikes of  branes and antibranes are
well separated, wormhole would not be  formed and there is no  channel
for flowing energy from extra dimensions to our universe.   On the other hand,
when two branes are close to each other and connected by a
wormhole, the bulk degrees of freedom  gain to large values.

At this stage, we can calculate the relation between some of
cosmological parameters like Hubble parametter and energy density
and some properties of BIon. For example,  the number of degrees
of freedom on the spherical surface of apparent horizon with
radius $r_{A}$ is proportional to its area and is given by:
\begin{eqnarray}
&& N_{sur} = \frac{4\pi r_{A}^{2}}{L_{P}^{2}}\label{QI11}
\end{eqnarray}
where $r_{A} = \sqrt{H^{2} + \frac{\bar{k}}{a^{2}}}$ is  the
apparent horizon radius for the FRW Universe, $H=
\frac{\dot{a}}{a}$ is the Hubble parameter and a is the scale
factor. Using this equation, we can estimate the Hubble parameter
for flat universe:
\begin{eqnarray}
&& H \simeq \frac{18\pi k^{4}N^{12}
T_{F1}^{14}}{T_{D3}^{14}\sigma_{0}^{8}} T^{12} + \frac{8\pi
k^{2}N^{10} T_{F1}^{12}}{T_{D3}^{12}\sigma_{0}^{4}}
T^{10}\label{QII11}
\end{eqnarray}
This equation has some interesting results which can be used to
explain the reasons for occurrence of expansion in present era of
universe. According to these calculations, the expansion of
universe is controlled by the number of branes, F-string and brane
tensions, the location of throat of wormhole and temperature of
BIon.

On the other hand,  using the Friedmann equation of the flat FRW
Universe, we can calculate the universe energy density:
\begin{eqnarray}
&& \rho = \frac{3}{8\pi L_{P}^{2}}H^{2} \simeq \frac{27\pi
k^{8}N^{24} T_{F1}^{28}}{4L_{P}^{2}T_{D3}^{28}\sigma_{0}^{16}}
T^{24} + \frac{3\pi k^{4}N^{20}
T_{F1}^{24}}{L_{P}^{2}T_{D3}^{24}\sigma_{0}^{8}}
T^{20}\label{QIII11}
\end{eqnarray}
As is obvious from  this equation, the energy density depends on
the temperature of BIon and any enhancement or decrease in this
density can be a signature of some interactions between two
universes in extra dimension. The reason of this is that with the  moving of  two
branes in the directions to  each other, radiation of energy from wormhole raises
and obtains  the large values near the colliding point.

Now, the main question arises that what is the fate of universe at
the end of expansion? To search  this question, we match  the
  finite temperature BIon and black F-string at
corresponding point\cite{f9}.  We have the supergravity solution for k
coincident non-extremal black F-strings lying along the z
direction as
\begin{eqnarray}
&& ds^{2} = H^{-1}(-f dt^{2} + dz^{2})+ f^{-1}dr^{2} + r^{2}d\Omega_{7}^{2}\nonumber\\
&& e^{2\phi} = H^{-1},\: B_{0} = H^{-1}-1,\nonumber\\
&& H = 1 +
\frac{r_{0}^{6}sinh^{2}\alpha}{r^{6}},\:f=1-\frac{r_{0}^{6}}{r^{6}}
\label{Q12}
\end{eqnarray}
here written in the string frame. From this the mass density along
the z direction can be found using Ref.\cite{f10}:
\begin{eqnarray}
&& \frac{dM_{F1}}{dz} = \frac{3^{5}T_{D3}^{2}(1 + 6cosh^{2}\alpha)}{2^{7}\pi^{3}T^{6}cosh^{6}\alpha}\nonumber\\
&&k^{2} = \frac{3^{12}T_{D3}^{4}(-1 +
cosh^{2}\alpha)}{2^{12}\pi^{6}T_{F1}^{2}T^{12}cosh^{10}\alpha},\:T_{F1}
= \frac{1}{2\pi l_{s}^{2}}\label{Q13}
\end{eqnarray}
For small temperatures one can
expand the mass density as follows:
\begin{eqnarray}
&& \frac{dM_{F1}}{dz} = T_{F1}k +
\frac{16(T_{F1}k\pi)^{3/2}T^{3}}{81T_{D3}}+
\frac{40T_{F1}^{2}k^{2}\pi^{3}T^{6}}{729T_{D3}^{2}}\label{Q14}
\end{eqnarray}
On the other hand, for small temperature BIon, we have \cite{f9}:
\begin{eqnarray}
&& \frac{dM_{BIon}}{dz} = T_{F1}k + \frac{3\pi T_{F1}^{2}k^{2} T^{4}}{32T_{D3}^{2}\sigma_{0}^{2}}+
 \frac{7\pi^{2} T_{F1}^{3}k^{3} T^{8}}{512T_{D3}^{4}\sigma_{0}^{4}}\label{Q15}
\end{eqnarray}
If one compares  the mass densities for BIon  to the mass density for the
F-strings, one will see that   the thermal D3-F1
configuration at $\sigma = \sigma_{0}$    behaves like k
F-strings and  $\sigma_{0}$ should have the following dependence on
the temperature as :
\begin{eqnarray}
&& \sigma_{0} =
\left(\frac{\sqrt{kT_{F1}}}{T_{D3}}\right)^{1/2}\sqrt{T}\left[C_{0} +
C_{1}\frac{\sqrt{kT_{F1}}}{T_{D3}}T^{3}\right]\label{Q16}
\end{eqnarray}
At this point, we can obtain the the degrees of  the surface,
degrees of freedom on the holographic horizon and the bulk degrees
of freedom inside the universe as  ( here, we have assumed
infinitely thin shell  wormhole that connect these branes ) :
\begin{eqnarray}
&& N_{sur} - N_{bulk} \simeq
\int_{\sigma_{0}}^{\sigma_{0}+\epsilon} d\sigma
\frac{dM_{BIon}}{dz} \rightarrow \nonumber
\\&& N_{sur} - N_{bulk} = T_{F1}k\epsilon + \frac{3\pi T_{F1}^{2}k^{2} T^{4}\epsilon}{32T_{D3}^{2}\sigma_{0}(\sigma_{0}+\epsilon)}+
 \frac{(7\pi^{2} T_{F1}^{3}k^{3} T^{8})(\epsilon^{2}-2\sigma_{0}\epsilon)}{512T_{D3}^{4}\sigma_{0}^{2}(\sigma_{0}+\epsilon)^{2}}\nonumber
\\&& \lim_{\epsilon\rightarrow 0} {[N_{sur} - N_{bulk}]} =0 \rightarrow\ \nonumber
\\&&N_{sur} = N_{bulk}\rightarrow \nonumber
\\&& N_{sur} + N_{bulk} = 2 N_{sur}= N_{black F-string} \label{Q17}
\end{eqnarray}
This equation indicates that the bulk degrees of freedom inside
the universe will be equal to the degrees of freedom of black
F-string at the end of universe expansion. This means that
universe evolves by the the phenomenological events and wormholes
in extra dimension and ends up in one black F-string.

\section{The Padmanabhan idea in eleven dimensional M2-M5 BIonic system}\label{o2}
In this section we will enter the effects of evolution in M2-M5
BIonic system on the surface and the bulk degrees of freedom in
FRW model of cosmology. We will show that contrary to previous
section, universe expands and ends up in black M2-brane.

To describe the BIon we specialize to an embedding of the M5-brane
world volume in 11D Minkowski space-time with metric
\cite{f11,f12}:
\begin{eqnarray}
&& ds^{2} = -dt^{2} +  (dx^{1})^{2} + dr^{2} +
r^{2}d\Omega_{3}^{2} + \sum_{i=6}^{10}dx_{i}^{2}. \label{Q18}
\end{eqnarray}
without background fluxes. Using the standard angular coordinates
$(\psi,\phi,\omega)$ to express the round three-sphere metric
\begin{eqnarray}
&& d\Omega_{3}^{2} = -d\psi^{2} +  sin^{2}\psi ( d\phi^{2} +
sin^{2}\phi d\omega^{2}). \label{Q19}
\end{eqnarray}
We choose the static gauge:
\begin{eqnarray}
&&t(\sigma^{a}) = \sigma^{1},\,x^{1}(\sigma^{1}) =
\sigma^{1},\,r(\sigma^{a})=\sigma^{2}\equiv \sigma  \nonumber \\&&
\psi(\sigma^{a}) = \sigma^{3} ,\, \phi(\sigma^{a}) = \sigma^{4},\,
\omega(\sigma^{a}) = \sigma^{5},\, x^{6}(\sigma^{a}) =
z(\sigma)\label{Q20}
\end{eqnarray}
 With this ansatz the
induced metric on the effective fivebrane world volume is
\begin{eqnarray}
\gamma_{ab}d\sigma^{a}d\sigma^{b} = -(d\sigma^{0})^{2} +
(d\sigma^{1})^{2} + (1 + z'(\sigma)^{2})d\sigma^{2} +
\sigma^{2}(-d\psi^{2} +  sin^{2}\psi ( d\phi^{2} + sin^{2}\phi
d\omega^{2})) \label{Q21}
\end{eqnarray}

 We impose the two boundary conditions that $z(\sigma)\rightarrow 0$ for $\sigma\rightarrow \infty$ and $z'(\sigma)\rightarrow -\infty$ for $\sigma\rightarrow \sigma_{0}$, where $\sigma_{0}$ is the minimal two-sphere radius of the
configuration. After some algebra, we obtain \cite{f11,f12}:
\begin{eqnarray}
z_{\pm}(\sigma)= \int_{\sigma}^{\infty} ds
\left(\frac{F_{\pm}(s)^{2}}{F_{\pm}(\sigma_{0})^{2}}-1\right)^{-\frac{1}{2}}
\label{Q22}
\end{eqnarray}
In finite temperature BIon, $F(\sigma)$ is given by
\begin{eqnarray}
F_{\pm}(\sigma) = \sigma^{3}\left(\frac{1 + \frac{k^{2}}{\sigma^{6}}}{1
\pm \sqrt{1 - \frac{4q_{5}^{2}}{\beta^{6}}(1 +
\frac{k^{2}}{\sigma^{6}})}}\right)^{3/2}\left(-2 +
\frac{3\beta^{6}}{2q_{5}^{2}}\frac{1 \pm \sqrt{1 -
\frac{4q_{5}^{2}}{\beta^{6}}(1 + \frac{k^{2}}{\sigma^{6}})}}{1 +
\frac{k^{2}}{\sigma^{6}}}\right) \label{Q23}
\end{eqnarray}
where
\begin{eqnarray}
\beta = \frac{3}{4\pi T} ,\, q_{2} =
\sigma^{3}\frac{r_{0}^{3}}{2}sin\theta sinh2\alpha  ,\, q_{5} =
\frac{r_{0}^{3}}{2}cos\theta sinh2\alpha\label{Q24}
\end{eqnarray}
with the definitions:
\begin{eqnarray}
&&cosh\alpha_{\pm} = \frac{\beta^{3}}{\sqrt{2}q_{5}}\frac{\sqrt{1
\pm \sqrt{1 - \frac{4q_{5}^{2}}{\beta^{6}}(1 +
\frac{k^{2}}{\sigma^{6}})}}}{\sqrt{1 +
\frac{k^{2}}{\sigma^{6}}}}\nonumber \\&& r_{0,\pm} =
\frac{\sqrt{2}q_{5}}{\beta^{2}}\frac{\sqrt{1 +
\frac{k^{2}}{\sigma^{6}}}}{\sqrt{1 \pm \sqrt{1 -
\frac{4q_{5}^{2}}{\beta^{6}}(1 + \frac{k^{2}}{\sigma^{6}})}}}
\nonumber \\&& tan\theta = \frac{k}{\sigma^{3}},\,q_{2} = k q_{5}
= -4\pi \frac{N_{2}}{N_{5}}l_{p}^{3}
 \label{Q25}
\end{eqnarray}
In above equation, $N_{2}$ and $N_{5}$ are the number of M2 and
M5-branes and $q_{2}$ and $q_{5}$ are the charges of M2 and
M5-branes respectively and T is temperature of BIon.  Attaching a
mirror solution to Eq. (\ref{Q25}), we construct thin shell  wormhole
configuration. The separation distance $\Delta = 2z(\sigma_{0})$
between the N M5-branes and N anti M5-branes for a given
brane-antibrane wormhole configuration is obtained by:
\begin{eqnarray}
\Delta = 2z(\sigma_{0})= 2\int_{\sigma_{0}}^{\infty}
ds\left(\frac{F(s)^{2}}{F(\sigma_{0})^{2}}-1\right)^{-\frac{1}{2}}
\label{Q26}
\end{eqnarray}

  Let us now construct the Padmanabhan idea in M2-M5 BIon. Similar to previous section, we
need to compute the contribution of the M2-M5 system to the
degrees of the surface degrees of freedom on the holographic
horizon and the bulk degrees of freedom inside the universe. To
this end, we write the following relations between these degrees
of freedom and the entropy of M2-M5 and also, the mass density
along the transverse direction,
 \begin{eqnarray}
&&N_{sur} + N_{bulk} = N_{M2-M5} = N(M5-brane) + N(anti-M5-brane)
+ N(M2-brane) \nonumber \\&&\simeq
4L_{P}^{2}S_{M2-M5}=\frac{\Omega_{(3)}\Omega_{(4)}}{16\pi G}\int
d\sigma\frac{F(\sigma)}{F(\sigma_{0})}\beta^{4}\sigma^{3}\frac{1}{cosh^{3}\alpha}
\nonumber \\&& N_{sur} - N_{bulk} \simeq \int d\sigma
\frac{dM_{M2-M5}}{dz}=\frac{\Omega_{(3)}\Omega_{(4)}}{16\pi G}\int
d\sigma
\frac{F(\sigma)}{F(\sigma_{0})}\beta^{3}\sigma^{3}\frac{3cosh^{2}\alpha
+ 1}{cosh^{3}\alpha} \label{Q27}
\end{eqnarray}
Solving these equations simultaneously, we obtain:
\begin{eqnarray}
&&N_{sur} \simeq \frac{\Omega_{(3)}\Omega_{(4)}}{16\pi G}\int
d\sigma\frac{F(\sigma)}{F(\sigma_{0})}\beta^{4}\sigma^{3}\frac{1}{cosh^{3}\alpha}
+ \frac{\Omega_{(3)}\Omega_{(4)}}{16\pi G}\int d\sigma
\frac{F(\sigma)}{F(\sigma_{0})}\beta^{3}\sigma^{3}\frac{3cosh^{2}\alpha
+ 1}{cosh^{3}\alpha} \nonumber \\&&  N_{bulk} \simeq
\frac{\Omega_{(3)}\Omega_{(4)}}{16\pi G}\int
d\sigma\frac{F(\sigma)}{F(\sigma_{0})}\beta^{4}\sigma^{3}\frac{1}{cosh^{3}\alpha}
- \frac{\Omega_{(3)}\Omega_{(4)}}{16\pi G}\int d\sigma
\frac{F(\sigma)}{F(\sigma_{0})}\beta^{3}\sigma^{3}\frac{3cosh^{2}\alpha
+ 1}{cosh^{3}\alpha} \label{Q28}
\end{eqnarray}
This equation implies that any increase or decrease in the bulk or
surface degrees of freedom can be a signature of some evolutions
in M2-M5 brane. Also, similar to the results of previous section,
with decreasing the distance between two branes, wormhole will be
formed and as a result, the bulk degrees of freedom increase.

Similar to previous section, we can obtain Hubble parameter and
energy density in terms of the number and charges of M2 and M5
branes.  Using equation (\ref{QI11}), we can estimate the Hubble
parameter for flat universe:
\begin{eqnarray}
&& H \simeq \frac{16\pi G k^{4}
q_{2}^{6}}{3\Omega_{(3)}\Omega_{(4)}q_{5}^{6}\sigma_{0}^{3}} T^{3}
+ \frac{8\pi G k^{2}
q_{2}^{4}}{\Omega_{(3)}\Omega_{(4)}q_{5}^{4}\sigma_{0}^{3}}
T^{2}\label{QI28}
\end{eqnarray}
This equation indicates that Hubble parameter depends on the
charges of M2 and M5 branes and also temperature of M2-M5 system.
Comparing this equation to (\ref{QII11}), we find that the Hubble
parameter in M2-M5 system is less sensitive to temperature respect
to BIonic system.

 In addition,  using the Friedmann equation of
the flat FRW Universe, we can get the universe energy density:
\begin{eqnarray}
&& \rho = \frac{3}{8\pi L_{P}^{2}}H^{2} \simeq \frac{384\pi G^{2}
k^{8}
q_{2}^{12}}{3L_{P}^{2}\Omega_{(3)}^{2}\Omega_{(4)}^{2}q_{5}^{12}\sigma_{0}^{6}}
T^{6} + \frac{24\pi G^{2} k^{4}
q_{2}^{8}}{L_{P}^{2}\Omega_{(3)}^{2}\Omega_{(4)}^{2}q_{5}^{8}\sigma_{0}^{6}}
T^{4} \label{QII28}
\end{eqnarray}
As can be seen from this equation, the energy density is related
to the charges of M-branes and the location of throat of wormhole
in M2-M5 system. When two M5-branes becomes close to each other
and thin shell wormhole is formed, energy flows from extra dimensions to our
universe and energy density increases.

Now, we try to know  the fate of universe in M2-M5 system. To
search  this question, we demand the matching of finite
temperature BIon and black M2-M5 at corresponding point\cite{f12}.
The M2 brane has the same charge $Q_{2}$ and the same temperature
T as the M2-M5 system. A perturbative expansion of the tension
around the extremal limit give \cite{f12}:

\begin{eqnarray}
\frac{M_{M2-brane}}{L_{x^{1}}L_{z}}= Q_{2}\left(1+
\frac{\sqrt{q_{2}}}{3\sqrt{2}\beta^{3}}+\frac{5q_{2}}{2^{6}\beta^{6}}\right)
\label{Q29}
\end{eqnarray}

On the other hand, for small temperature M2-M5 system, we have
\cite{f12}:
\begin{eqnarray}
&& \frac{dM_{BIon}}{L_{x^{1}}dz}=
Q_{2} \left(\sqrt{1+\frac{\sigma^{6}}{k^{2}}} +
\frac{5q_{2}^{2}}{6\beta^{6}}\frac{\left(1+\frac{\sigma^{6}}{k^{2}}\right)^{3/2}}{\sigma_{0}^{6}}
+
\frac{11q_{2}^{4}}{8\beta^{12}}\frac{\left(1+\frac{\sigma^{6}}{k^{2}}\right)^{5/2}}{\sigma_{0}^{12}}
\right)\label{Q30}
\end{eqnarray}
 
We now compare the mass densities for M2-M5 system to the mass density
for black M2-brane and note that    the  thermal BIon at
$\sigma = \sigma_{0}$  behaves like black M2-brane. $\sigma_{0}$ 
depends  on the temperature as
\cite{f12}:
\begin{eqnarray}
&& \sigma_{0} = \frac{q_{2}^{1/4}}{\beta^{1/2}}\left(1.234 -
0.068\frac{q_{2}^{1/2}}{\beta^{3}}\right) \label{Q31}
\end{eqnarray}
At this point, we can obtain the the degrees of  the surface
degrees of freedom on the holographic horizon and the bulk degrees
of freedom inside the universe as  (as before, we have
assumed infinitely thin shell  wormhole that connect these branes
) ::
\begin{eqnarray}
&& N_{sur} - N_{bulk} \simeq
\int_{\sigma_{0}}^{\sigma_{0}+\epsilon} d\sigma
\frac{dM_{M2-M5}}{dz}\rightarrow  \nonumber
\\&&N_{sur} - N_{bulk}\simeq Q_{2}\left(\frac{\epsilon^{7/2}}{\sigma_{0}^{7}k^{2}} +
\frac{15q_{2}^{2}}{84\beta^{6}}\frac{\epsilon^{21/2}}{k^{2}\sigma_{0}^{13}}
+
\frac{55q_{2}^{4}}{112\beta^{12}}\frac{\epsilon^{35/2}}{k^{2}\sigma_{0}^{12}}
\right) \nonumber
\\&& \lim_{\epsilon\rightarrow 0} {[N_{sur} - N_{bulk}]} =0 \rightarrow\  \nonumber
\\&& N_{sur} = N_{bulk}\rightarrow \nonumber
\\&& N_{sur} + N_{bulk} = 2N_{sur} =N_{black M2-brane} \label{Q32}
\end{eqnarray}
This equation indicates that the bulk degrees of freedom inside
the universe will be equal to the degrees of freedom of black
M2-brane at the end of universe expansion. This result is in
contrary with the results of previous section that showed the
final state of universe will be a black F-string.

\section{Summary and Discussion} \label{sum}
In this research, we  construct Padmanabhan in thermal BIon and
discuss that the surface degrees of freedom on the holographic
horizon and the bulk degrees of freedom inside the universe depend
on the temperature of BIon, number of branes and the distance
between branes. For large separation distance of  branes and antibranes  
as well as  the brane's spike is far from the antibrane's spike, the role
of wormhole is ignorable,
 however,  when  the spikes of brane and antibrane
comes closer to  each other, one  thin shell  wormhole would be
formed. In this condition, there is many channels for flowing
energy from extra dimensions into our universe, the bulk degrees
of freedom increase and tend to the degrees of freedom of one
black F-string in string theory or black M2-brane in M-theory.

\section*{Acknowledgments}
\noindent FR   wishes to thank the authorities of the
Inter-University Centre for Astronomy and Astrophysics, Pune,
India for providing the Visiting Associateship. IHS is also thankful to  DST, Govt.
of India for providing financial support.

 \end{document}